\documentclass[supplement]{ptptex}

\usepackage{graphicx}
\usepackage{bm}

\title{Nematic order in the multiple-spin exchange model \\
on the triangular lattice}

\author{Tsutomu \textsc{Momoi}$^{1)}$ and Nic
\textsc{Shannon}$^{2)}$
}

\inst{${}^1$Condensed Matter Theory Laboratory, RIKEN, Wako,
Saitama 351-0198, Japan\\
${}^2$Max--Planck--Institut f{\"u}r Physik komplexer Systeme,
  N{\"o}thnitzer Str. 38, 01187 Dresden, Germany
  }

\recdate{\hspace*{4cm}}

\abst{We discuss the possibility of finding $n$-type nematic order
in the vicinity of a FM phase in the multiple--spin exchange model
on the triangular lattice. We study this problem both from
$S=\infty$ (classical) and $S=1/2$ (quantum) limits. }

\begin{document}

\maketitle

The possibility of realizing a quantum spin liquid in a frustrated
magnet with antiferromagnetic interactions has been widely
discussed. Here we consider the nature of a quantum spin liquid
appearing in the vicinity of a {\it ferromagnetic} phase. Our
motivation is the gapless spin liquid observed in two-dimensional
(2D) solid $^3$He on graphite\cite{Fukuyama,Ishimoto}, very close
to a ferromagnetic phase. This system is believed to be well
described by a multiple--spin exchange (MSE) model on the
triangular lattice:
\begin{equation}\label{Hamiltonian}
{\cal H}= J \sum_{\rm N.N.} P_2 + K \sum_p (P_4+P_4^{-1}) -H
\sum_i S^z_i,
\end{equation}
where $P_n$ denotes the cyclic permutation operator of n spins.
The first summation runs over all nearest neighbor pairs and the
second one all four-site minimal diamond plaquettes. In 2D solid
$^3$He, the effective two--spin exchange coupling $J<0$ is {\it
ferromagnetic} (FM) while the four-spin cyclic exchange $K>0$ is
antiferromagnetic (AF). An exact--diagonalization study of
finite--size systems\cite{MisguichBLW} concluded that the FM phase
for small $K$ evolves into a state with zero total spin for large
values of $K$ and that, if $K$ is quite large, the ground state is
spin liquid with no long--range magnetic order and a spin gap of
order $|J|$. How such a spin liquid might form out of a FM state,
and what the nature of the spin liquid should be in the vicinity
of the FM phase, remain open problems.

Here, we present evidence that the first instability of the FM
state with increasing $K$ is against a gapless spin liquid state
with nematic order. We consider the $n$-type spin nematic order
parameter, defined by
\begin{equation}
    \label{eqn:n-type}
{\cal O }^{\alpha\beta}(\boldsymbol{r}_i, \boldsymbol{r}_j)
= \frac{1}{2}(S^\alpha_i S^\beta_j+S^\beta_i S^\alpha_j) - {1
\over 3} \delta^{\alpha\beta}\langle \boldsymbol{S}_i \cdot
\boldsymbol{S}_j \rangle.
\end{equation}
This type of spin nematic order has been seen in the $S=1$
bilinear--biquadratic model \cite{HaradaK}.
However, to the best of our knowledge, it has never before
been found in a spin $S=1/2$ frustrated system.

Let us consider first the classical limit of the MSE model on the
triangular lattice.  A mean--field argument and its
extension\cite{MomoiSK} showed that a highly degenerate phase
appears in the wide parameter range $-1<K/J<-1/4$. This set
includes all parameters relevant to 2D solid $^3$He in the low
density limit, and extends up to the classical boundary of the FM
phase at $K/J = -1/4$. Among the many possible classical ground
states, collinear states are most likely to be favored by quantum
or thermal fluctuations and the two most characteristic collinear
ground states are the UUUD and UDDD states with exactly half the
saturation magnetization\cite{MomoiSK}. However, because it costs
no energy to add a straight domain wall between these states,  at
any finite temperature the system will gain entropy by breaking up
into stripe--like UUUD and UDDD domains. For this reason we do not
expect any long range spin--spin correlations to form in this
system, even in the limit $T \to 0$.

\begin{figure}[!tb]
     \includegraphics[width=65mm]{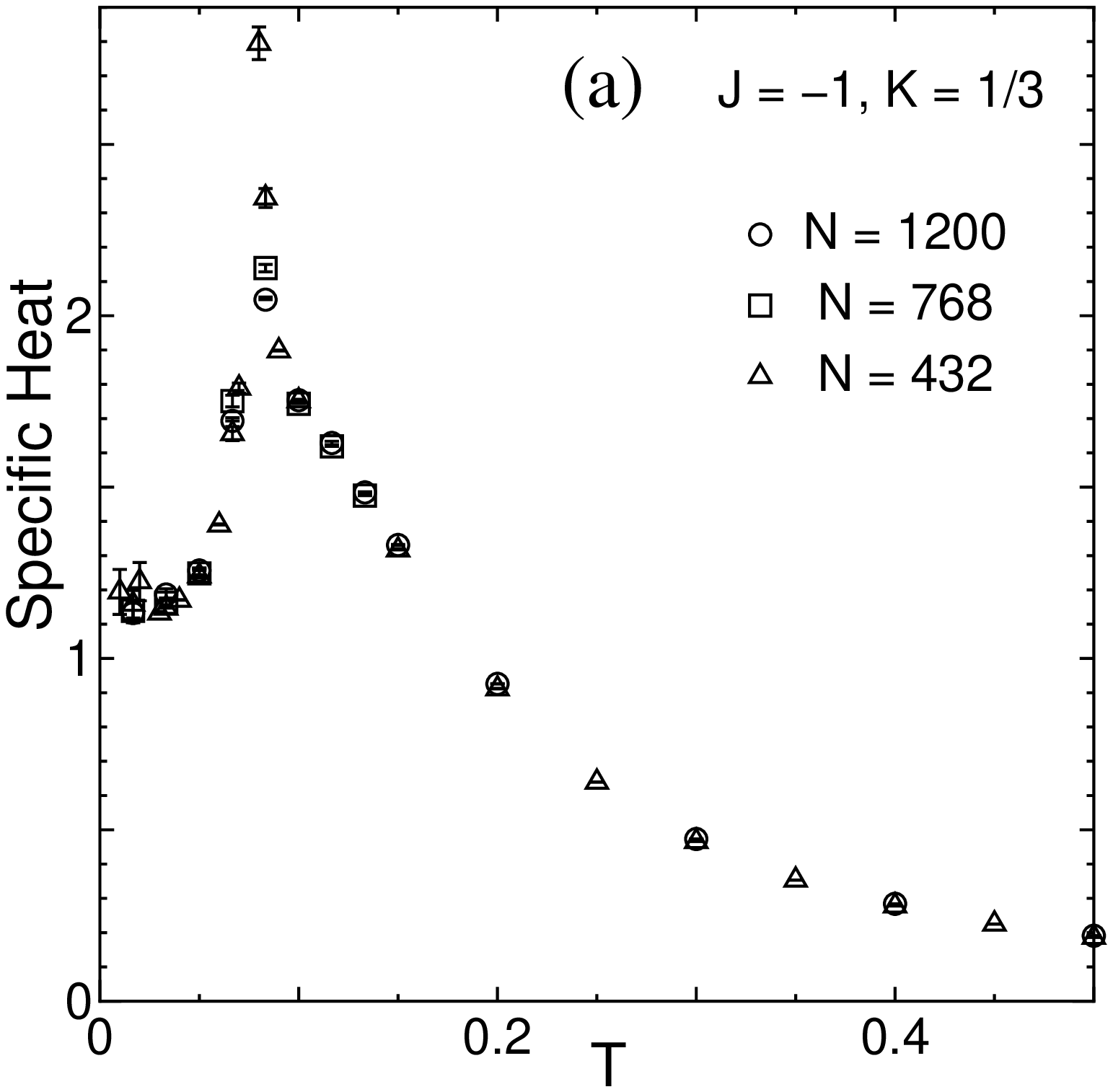}
 \hspace{5mm}
     \includegraphics[width=65mm]{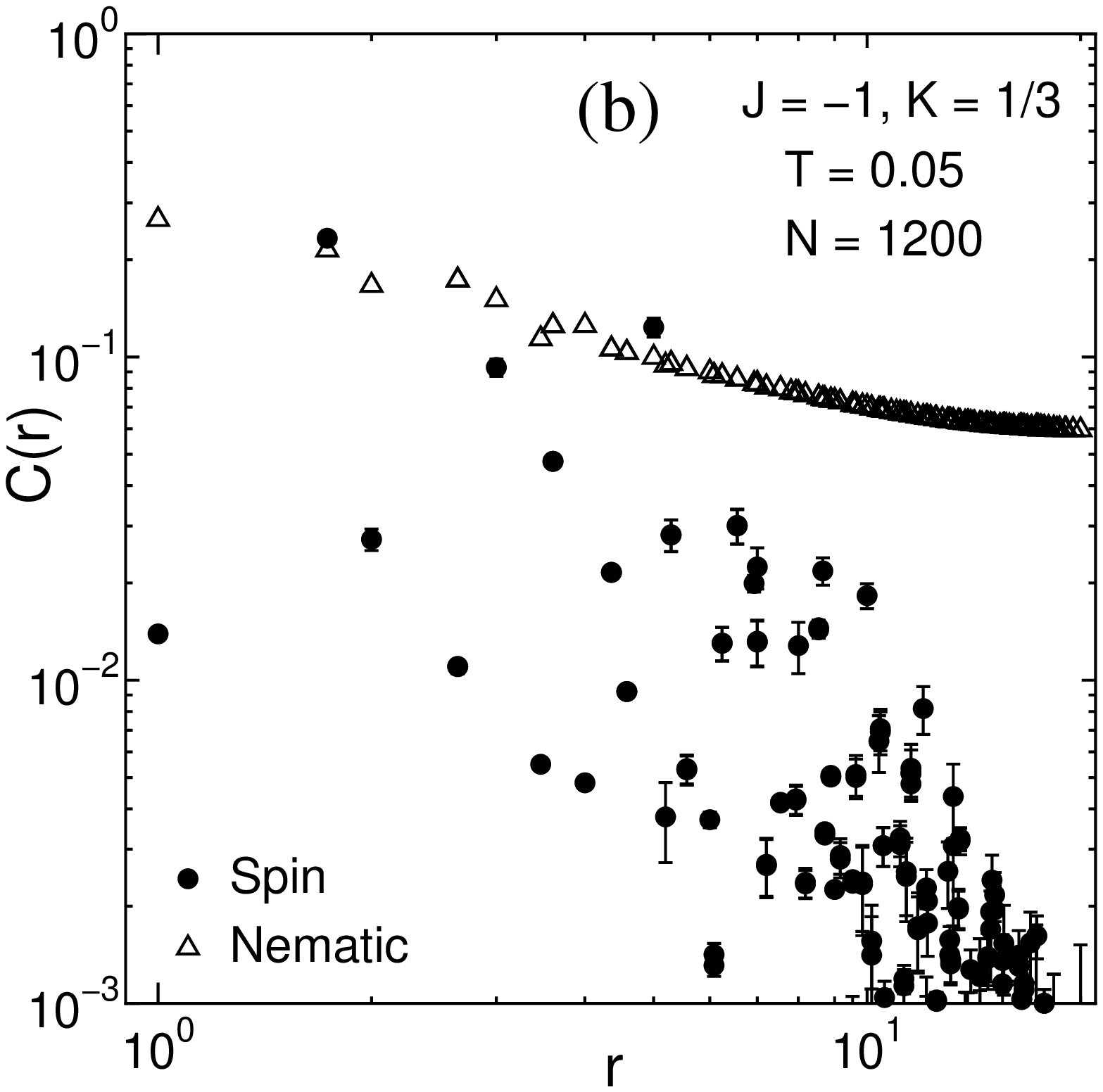}
 \caption{(a) Temperature dependence of specific heat in the classical MSE
 model for $J=-1$, $K=1/3$, $H=0$ with peak corresponding to the
 onset of nematic correlations.
 Results are shown for systems of  $N=432$, $768$ and $1200$ spins.
 (b) Decay of spin
 (circle) and nematic (triangle) correlations at $T=0.05$ for
 the classical MSE model with $J=-1$, $K=1/3$, $H=0$, and $N=1200$.   All
 temperatures are measured in units of $|J|$.}
 \label{fig:corr_both0.16T0.05L10}
 \end{figure}

However we {\it do} expect that the spins will remain
collinear, and that this will lead to long ranged spin nematic
correlations of the form ${\cal O }^{\alpha\beta}(\boldsymbol{r}_i,
\boldsymbol{r}_j)$, 
above.
To test this argument, we performed MC simulations of the classical
model and found that nematic correlations are enhanced at low
temperatures, while spin correlations decay quite rapidly. (See
Fig.\ \ref{fig:corr_both0.16T0.05L10}.)    The onset of nematic
correlations is signaled by a peak in the specific heat whose
height remains finite as the system size is increased.
Of course, true long ranged nematic order
cannot be achieved in 2D at any finite temperature
because the director (here, the axis of collinearity)
breaks a continuous symmetry.

\begin{figure}[!tb]
 \begin{center}
     \includegraphics[width=85mm]{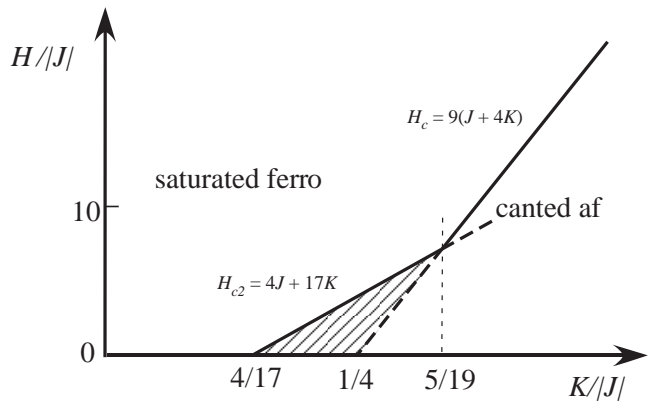}
 \caption{Saturation fields: the parameter $K/|J|$ vs.\ the one-- and
 two--magnon saturation fields $H_{c}$ and $H_{c2}$.   In the shaded
 region, the bi-magnon bound states have a negative energy relative to
 the saturated FM state, while individual magnons have a
 positive energy with a finite positive stiffness. Thus the
 partially polarized spin state in the shade should show BEC of
 bound bi--magnons.} \label{fig:PD}
 \end{center}
 \end{figure}

We now turn to the experimentally relevant quantum limit, i.e. the $S=1/2$
MSE model, and consider the nature of the first instability of the FM
as $K$ is increased.   In the absence of magnetic field, the entire one
magnon spectrum
\begin{eqnarray}
\epsilon(\boldsymbol{k}) = -2(J+4K) \{3 - \cos
(\boldsymbol{k}\cdot \boldsymbol{e}_1) -\cos (\boldsymbol{k}\cdot
\boldsymbol{e}_2) -\cos [\boldsymbol{k}\cdot
(\boldsymbol{e}_1-\boldsymbol{e}_2)]\}
\end{eqnarray}
{\it vanishes} at the classical phase boundary $K/J=-1/4$,
reflecting the extensive degeneracy of the classical disordered
phase beyond it. For large values of applied magnetic field, this
degeneracy is lifted and the first instability of the saturated FM
state is against a three--sublattice canted AF state. However
for smaller applied fields, the system can gain kinetic energy by
allowing pairs of flipped spins to move together through repeated
use of the cyclic exchange process. This prompts us to consider
the following trial bi--magnon bound state:
\begin{equation}
a(\boldsymbol{k})|1(\boldsymbol{k})\rangle +
b(\boldsymbol{k})|2(\boldsymbol{k})\rangle
\end{equation}
with
\begin{eqnarray} |1(\boldsymbol{k})\rangle &=& N^{-1/2}\sum_i
S_i^- S_{i+e_1}^-
e^{i\boldsymbol{k}\cdot(\boldsymbol{r}_i+\boldsymbol{e}_1/2)}
|PF\rangle,\nonumber \\
|2(\boldsymbol{k})\rangle &=& N^{-1/2}\sum_i S_i^- S_{i+e_2}^-
e^{i\boldsymbol{k}\cdot(\boldsymbol{r}_i+\boldsymbol{e}_2/2)}
|PF\rangle.\nonumber
\end{eqnarray}
Despite the fact that the static interaction between pairs of
magnons is repulsive, for $K/J<-4/17$, there exist trial solutions
with a negative energy relative to the saturated FM state, i.e.
bi--magnon bound states below the FM continuum. Thus the first
instability of the saturated FM state in zero field is against the
spontaneous creation of bound bi--magnon pairs. Extending this
calculation to finite magnetic field, we obtain the phase diagram
shown in Fig.\ \ref{fig:PD}.

In order to better understand the unsaturated phase bordering on
the FM phase let us consider the dynamics of bound bi-magnons. We
know that pairs of flipped spins have repulsive interactions and a
definite (positive) hopping amplitude.   From this we can write
down an effective Hamiltonian for bound bi--magnons, which has a
form similar to an AF XXZ model. We therefore anticipate that the
density of bound bi-magnons increases continuously upon decreasing
the magnetic field from the critical value $H_{c2}$, and that
there is a second order quantum phase transition from the
saturated to the unsaturated state.

To be certain that bound bi--magnons are the elementary particles
of the unsaturated state, we need to rule out the existence of
larger magnon bound states.    Cooperative motion of more than
three flipped spins can occur in high order perturbation theory in
$K$.  However the repulsive interaction between magnons is first
order in $K$, so we do not, in general, expect larger bound states
to be stable. From these arguments we conclude that the
unsaturated states close to the saturation field are characterized
by the Bose--Einstein condensation (BEC) of a low density of bound
bi-magnons.

Finally we consider the nature of this bound bi--magnon
condensate.     The magnon pairing operator is given by
\begin{equation}
{\cal O}_{\rm pair} (\boldsymbol{r}_i,\boldsymbol{r}_j) = s^-_i
s^-_j.
\end{equation}
This is related to the nematic operators through
\begin{align}
{\cal O}_{\rm pair} = s_i^- s_j^- = {\cal O }_{N}^{xx} - {\cal O
}_{N}^{yy} - 2i{\cal O }_{N}^{xy}, \label{eq:O_pair_nematic}
\end{align}
where we have omitted site indices. The real part of the pairing
operators corresponds to the nematic operator ${\cal O }_{N}^{xx}
- {\cal O }_{N}^{yy}$ and the imaginary part to $-2{\cal O
}_{N}^{xy}$.  These two components give nematic order parameters
for the systems under the magnetic field.  Thus, the BEC of bound
bi--magnons is equivalent to the emergence of long--ranged
spin nematic order.

Our conclusion is that the MSE model on the triangular lattice in the
vicinity of the saturated FM phase shows long ranged
spin nematic order in one of two channels
\begin{eqnarray}
\lim_{|r|\rightarrow\infty}\langle [{\cal
O}^{xx}(\boldsymbol{0},\boldsymbol{e}_1) - {\cal
O}^{yy}(\boldsymbol{0},\boldsymbol{e}_1)] &&[{\cal
O}^{xx}(\boldsymbol{r}_i, \boldsymbol{r}_i + \boldsymbol{e}_1) -
{\cal O}^{yy}(\boldsymbol{r}_i, \boldsymbol{r}_i
+ \boldsymbol{e}_1)] \rangle \nonumber\\
&& \sim c_1 (1-m/m_{sat})
\end{eqnarray}
or
\begin{eqnarray}
\lim_{|r|\rightarrow\infty} 4 \langle {\cal
O}^{xy}(\boldsymbol{0}, \boldsymbol{e}_1) {\cal
O}^{xy}(\boldsymbol{r}_i, \boldsymbol{r}_i+\boldsymbol{e}_1)
\rangle \sim c_1 (1-m/m_{sat}),
\end{eqnarray}
where $m$ $(m_{sat})$ denotes the (saturated) magnetization and
$c_1$ a finite constant.  In these states, spin correlations decay
exponentially even at zero temperature, making them as true
spin--liquids. The rate at which the spin correlations decay is
set by the binding energy of bound bi--magnon pairs, which
vanishes at the boundary with the saturated FM phase.

Correlated hopping processes\cite{MomoiTa} lead to a similar BEC
of bound bi--magnons in the low-magnetization regime of the 2D
Shastry--Sutherland model\cite{MomoiTb} .   This can also be
regarded as a spin nematic state. However, nematic order may be
difficult to observe in the ``Shastry--Sutherland'' compound
SrCu$_2$(BO$_3$)$_2$ because of the presence of
Dzyaloshinsky--Moriya interactions which break the bi--magnon
pairs. The spin interactions in solid $^3$He, on the other hand,
are purely isotropic in nature.   We therefore expect that nematic
order {\it is} experimentally accessible for a range of densities
of 2D solid $^3$He in applied  magnetic field.

We will return to this issue, and present further analytic and
numerical results for nematic order neighboring ferromagnetism in
the MSE model, elsewhere \cite{MomoiSS}.

\section*{Acknowledgements}
N.S. would like to thank Michel Roger for many helpful discussions
about $^3$He  and multiple spin exchange.

\end{document}